\documentclass[preprint,usenatbib,twocolumn]{aastex631}
\usepackage{placeins}
\usepackage{newtxtext,newtxmath}
\usepackage[T1]{fontenc}
\usepackage{ae,aecompl}
\usepackage{fix-cm} 
\usepackage{graphicx}	
\usepackage{multirow}
\usepackage{xcolor}

\newcommand{\mi}[1]{\textsf{m12i}}
\newcommand{\mf}[1]{\textsf{m12f}}
\newcommand{\mm}[1]{\textsf{m12m}}
\newcommand{\Msol}[1]{M\textsubscript{\(\odot\)}}

\shorttitle{Galactic accelerations in simulations}

\begin{document}

\title{The imprint of dark matter on the Galactic acceleration field}

\author[0000-0002-8354-7356]{Arpit Arora}
\affiliation{Department of Physics and Astronomy, University of Pennsylvania, 209 S 33rd St., Philadelphia, PA 19104, USA
}
\author[0000-0003-3939-3297]{Robyn E. Sanderson}
\affiliation{Department of Physics and Astronomy, University of Pennsylvania, 209 S 33rd St., Philadelphia, PA 19104, USA
}
\author[0000-0001-6711-8140]{Sukanya Chakrabarti}
\affiliation{Department of Physics and Astronomy, University of Alabama in Huntsville, 301 North Sparkman Drive, Huntsville, AL 35816, USA}

\author[0000-0003-0603-8942]{Andrew Wetzel}
\affiliation{Department of Physics \& Astronomy, University of California, Davis, CA 95616, USA}

\author[0000-0002-7746-8993]{Thomas Donlon II}
\affiliation{Department of Physics and Astronomy, University of Alabama in Huntsville, 301 North Sparkman Drive, Huntsville, AL 35816, USA}

\author[0000-0003-1856-2151]{Danny Horta}
\affiliation{Center for Computational Astrophysics, Flatiron Institute, 162 5th Ave, New York, NY 10010, USA}

\author[0000-0003-3217-5967]{Sarah R. Loebman}
\affiliation{Department of Physics, University of California, Merced, 5200 Lake Road, Merced, CA 95343, USA}

\author[0000-0003-2806-1414]{Lina Necib}
\affiliation{Department of Physics and MIT Kavli Institute for Astrophysics and Space Research, 77 Massachusetts Avenue, Cambridge, MA 02139, USA}
\affiliation{The NSF AI Institute for Artificial Intelligence and Fundamental Interactions, 77 Massachusetts Avenue, Cambridge, MA 02139, USA}

\author[0000-0001-5636-3108]{Micah Oeur}
\affiliation{Department of Physics, University of California, Merced, 5200 Lake Road, Merced, CA 95343, USA}


\begin{abstract}
Measurements of the accelerations of stars enabled by time-series extreme-precision spectroscopic observations, pulsar timing, and eclipsing binary stars in the Solar Neighborhood offer insights into the mass distribution of the Milky Way that do not rely on traditional equilibrium modeling. Given the measured accelerations, we can determine a total mass density and infer the amount of dark matter (DM) by accounting for the mass in stars, gas, and dust. Leveraging FIRE-2 simulations of Milky Way-mass galaxies, we compare vertical acceleration profiles between cold DM (CDM) and self-interacting DM (SIDM) with constant cross-section of 1 cm$^2$g$^{-1}$ across three halos with diverse assembly histories. Notably, significant asymmetries in vertical acceleration profiles near the midplane at fixed radii are observed in both CDM and SIDM, particularly in halos recently affected by mergers with satellites of Sagittarius/SMC-like masses or greater. These asymmetries offer a unique window into exploring the merger history of a galaxy. We show that SIDM halos manifest a more oblate shape and consistently exhibit higher local stellar and DM densities and steeper vertical acceleration gradients, up to 10-30\% steeper near the Solar Neighborhood. However, similar magnitude changes can arise from azimuthal variations in the baryonic components at a fixed radius and external influences like mergers, making it difficult to distinguish between CDM and SIDM using acceleration measurements in a single galaxy. 
\end{abstract}



\section{Introduction} \label{sec:intro}

Dark matter (DM) constitutes approximately 85\% of the matter in the universe \citep{collaboration2020planck}, but its nature remains elusive. The standard cold dark matter (CDM) model, which assumes that DM particles are collisionless and non-interacting, has been successful in explaining large-scale structures in the Universe. However, there are several inconsistencies at small scales, such as the too-big-to-fail tension, core-cusp tension, planes-of-satellites, and others \citep[e.g.][]{flores1994observational, moore1994evidence, moore1999cold, bullock2017small, tulin2018dark, sales2022baryonic}. 

One unresolved issue is the ``diversity of rotation curves'', where the observed rotation curves for dwarf satellites and Milky Way-mass galaxies show a stunning diversity, with inferred inner DM profiles ranging from cored to NFW-like and even more concentrated than NFW profiles \citep[e.g.][]{oman2015unexpected, zavala2019diverse}. On the other hand, CDM simulations predict a steeply rising central density profile, and reproducing such diversity in inner DM profiles remains a challenge even with additions of baryonic components with active feedback \citep{walker2011method, santos2020baryonic, ebisu2022constraining}.  

This discrepancy has spurred exploration into alternative models, such as self-interacting dark matter (SIDM) \citep{carlson1992self, spergel2000sidm}, which allows DM particles to interact with each other through elastic collisions. Such interactions can lead to a transfer of kinetic energy from the hot outer region to the dense inner region, resulting in a more cored density profile \citep{spergel2000sidm, yoshida2000weakly, colin2002structure, gentile2004cored, tulin2018dark, salucci2019distribution}. Moreover, a large self-interaction cross-section ($\sigma/m \leq 10$ cm$^2$ g$^{-1}$) can induce a core-collapse phase that drives the DM mass towards the center and forms steeper inner DM profiles \citep{kahlhoefer2019diversity, zavala2019diverse, sameie2020self}. 

Recent studies have shown that SIDM can reproduce some observed properties of dwarf galaxies and Milky Way (MW)-mass galaxies better than CDM \citep{vogelsberger2012subhaloes, rocha2013cosmological, peter2013cosmological, vogelsberger2019evaporating}, such as their shapes \citep{sameie2018impact, sameie2021central, vargya2022shapes} and inner densities \citep{kaplinghat2016dark}. One effective test of DM involves measuring the DM mass distribution within a galaxy.

The most straightforward and least assumptive method for probing the mass distribution (including stars, gas and DM) in the Galaxy is through acceleration measurements of stars in the MW.
Recent advances in high-precision spectrographs \citep[e.g.,][]{schwab2016design, fischer2016state, Wright_2017,pepe2021espresso}, improving precision in pulsar timing data \citep[e.g.,][]{keith2013measurement, pennucci2019frequency, goncharov2021identifying}, and high-precision spectroscopic observations of eclipsing binary stars by space-based missions \citep{Helminiak2019} now enable direct measurement of Galactic accelerations in the Solar Neighborhood with multiple independent techniques \citep[e.g][]{quercellini2008mapping,Silverwood_Easther2019, chakrabarti2020toward, chakrabarti2021measurement, phillips2021milky,  Chakrabarti2022snowmass, chakrabarti2022eclipse}.

Acceleration measurements can be used to determine the shape of the MW potential in the Solar Neighborhood and constrain the Galactic mid-plane density \citep[Oort limit,][]{chakrabarti2020toward, chakrabarti2021measurement, donlon2024galactic}, from which we can determine the local DM density after accounting for the baryon budget. Notably, in comparison to alternative methodologies \citep[reviewed comprehensively by][]{pablo2020}, acceleration measurements require far fewer assumptions. Using Poisson's equation, given an acceleration field (i.e., the gradient of the potential), we can straightforwardly determine the local density. 

These properties of the MW depend on the DM model \citep{sameie2018impact, vargya2022shapes, sameie2021central}, indicating that measurement of the local dark matter density and the shape of the potential can provide discriminating power for constraining the nature of DM. Prior work on determinations of the Oort limit has focused on kinematic analysis \citep[e.g.][]{salucci2010dark, mckee2015stars, Schutz2018, Guoetal2020, de2021dark}, which models a snapshot in time of the positions and speeds of stars under simplifying assumptions of equilibrium and/or symmetry both across the midplane and axisymmetrically, which can lead to inaccurate inferences of Galactic parameters for a time-dependent potential \citep{Haines2019}. 

\cite{chakrabarti2020toward} demonstrated that perturbations from past mergers such as the Gaia-Sausage-Enceladus merger \citep{helmi2018merger, belokurov2018co}, the Sagittarius dwarf galaxy \citep{ibata1994dwarf, johnston1995disruption, newberg2002ghost}, the ongoing merger with the LMC \citep{besla2007magellanic, kallivayalil2013third} and other possible disturbances in the MW such as formation of a warped disk \citep{ostriker1989warped, evans1998lmc} or large planar disturbances in the outer gas disk of the Galaxy \citep{Chakrabarti_Blitz2009,Chakrabartietal2019} can induce asymmetries in the Galactic acceleration profile. 
These non-equilibrium effects are naturally taken into account while studying DM through direct measurement of Galactic accelerations, because extreme-precision time-series observations of stars provide the Galactic acceleration \emph{today} without assuming equilibrium or symmetry as in kinematic analyses. Moreover, kinematic analyses only yield the average acceleration for a bulk population of stars, which offers less constraining power.

Cosmological simulations with a realistic baryonic disk and assembly history can serve as a natural test laboratory for interrogating tools to measure local DM density \citep[e.g.][]{necib2019under, ou2024dark} and for interpreting Galactic acceleration observations \citep{loebman2012constraints, loebman2014milky}. Moreover, by examining these simulations across different DM models, we can effectively constrain the mass distribution and nature of DM.

In this paper, we use cosmological baryonic simulations with varying merger histories described in Sec.~\ref{sec:sims} to analyze their Galactic acceleration fields (Sec.~\ref{sec:gal_acc}) under different DM models. In Sec.~\ref{sec:oblt}, we compare the measured shape of the MW potential from pulsar timing with our simulations. Here, we compare to fundamental Galactic parameters determined from the time-rate of change of the orbital period of binary pulsars \citep{chakrabarti2021measurement, donlon2024galactic} as using the time-rate of change of the spin-period \citep{phillips2021milky} leads to large uncertainties due to the unknown effect of magnetic braking on the spin periods. Our conclusions are presented in Sec.~\ref{Sec:disc}.

\section{Simulations of MW-mass galaxies} \label{sec:sims}

We use three MW--mass galaxies using cosmological-baryonic simulations from the \textit{Latte} suite \citep{wetzel2023public} of FIRE-2 simulations with initial conditions derived from AGORA \citep{Agora2014} run with \texttt{GIZMO} \citep{hopkins2015new}. The halos are labelled m12f, m12i, and m12m are some of the most MW-like isolate disks in the suite and span a range of assembly histories. All the halos use identical baryonic FIRE-2 physics \citep{hopkins2018} and two distinct DM models: \textbf{CDM}, which employs cold-collisionless DM and \textbf{SIDM}, which adopts self-interacting DM with a cross-section of $\sigma/m = 1$ cm$^2$ g$^{-1}$.  The SIDM implementation in \texttt{GIZMO} (introduced in \citealt{rocha2013cosmological} and \citealt{peter2013cosmological})
 employs a Monte Carlo approach to determine the scattering probability for the nearest neighbors of each DM particle using a spline kernel with adaptive smoothing length \citep{monaghan1985refined}. It then assigns velocities isotropically to the scattered particles to conserve energy and momentum. The cross-section of $\sigma/m = 1$ cm$^2$ g$^{-1}$ for SIDM is often used for MW-mass galaxies to achieve a balance between addressing small-scale problems \citep{tulin2018dark} and maintaining agreement with larger-scale galactic and cluster observations \citep{randall2008constraints}.

These halos are all isolated systems with no massive companion halos within 4 Mpc, and have a virial mass of approximately $\sim 1-1.5 \times 10^{12}$ $\textrm{M}_{\odot}$. Each simulation uses an initial particle mass of $m_\mathrm{b} = 7100$ \Msol{} for stars and gas, and $m_\mathrm{DM} = 35000$ \Msol{} for DM with a minimum physical spatial resolution $\epsilon_\textrm{gas} = 1$ pc, $\epsilon_{\star} = 4$ pc, and $\epsilon_\textrm{DM} = 20$ pc. Also, the potential and acceleration values are tracked for every particle in the simulation.
 
The properties of the halos, including halo shapes and density profiles, along with a description of the simulation methods are detailed in \citet{sameie2021central} and \citet{vargya2022shapes}. These halos are part of a suite of simulations dedicated to exploring alternative DM models which are different from the fiducial FIRE-2 simulations presented in \citet{wetzel2023public}. In these simulations, any thermal energy to momentum for stellar winds energy conversion resulting from stellar mass-loss processes is neglected for the sub-resolution regions which leads to a lower (almost a factor of 2) stellar mass than our fiducial FIRE-2 simulations.

Notably, FIRE-2 CDM halos exhibit MW--like stellar-to-halo mass ratios \citep{hopkins2018}, stellar morphologies and kinematics \citep{ma2017structure, mccluskey2024disc}, and other properties such as gas fractions, scale radii, scale heights, and satellite populations \citep[e.g.][]{el2018gas, escala2018modelling, Sanderson2018, Samuel2021}, making them ideal for comparing with observational Galactic accelerations in the Solar Neighborhood. 

Additionally, the three halos have distinct mergers and assembly histories. The simulation m12i features a thick young disk with an intermediate formation epoch \citep{sand2020_prop} with a merger with first pericentric passage around 6 Gyr before the present day. m12f had a major merger with a pair of satellites similar in total mass to the progenitor of Sagittarius stream \citep{jiang2000orbit, niederste2010re, de2015star, gibbons2017tail} with the first pericentric passage about 3 Gyr before the present day \citep{arora2022stability, garavito2023co}. In contrast, m12m is the earliest forming, has a massive disk, and has no massive mergers in the last 7 Gyr of its evolution \citep{m12mform}. Table~\ref{tab:gal_prop} lists the total stellar mass within 10\% of the virial radius, where the cumulative density is 200 times the critical density of the Universe ($\textrm{M}_\star$) and the 3D stellar half-mass radius ($\textrm{r}_{1/2}$) for all of the galaxies. While the total virial mass of these systems is the same in the SIDM and CDM runs, the SIDM simulations have systematically higher stellar mass throughout the disk due to increased star formation rates at late times \citep{sameie2021central} and larger galaxy sizes, quantified by $\textrm{r}_{1/2}$, compared to the CDM simulations.

\subsection{The Solar Neighborhood} \label{sec:solar_neigh}

\begin{figure*}
\includegraphics[width=\textwidth,]{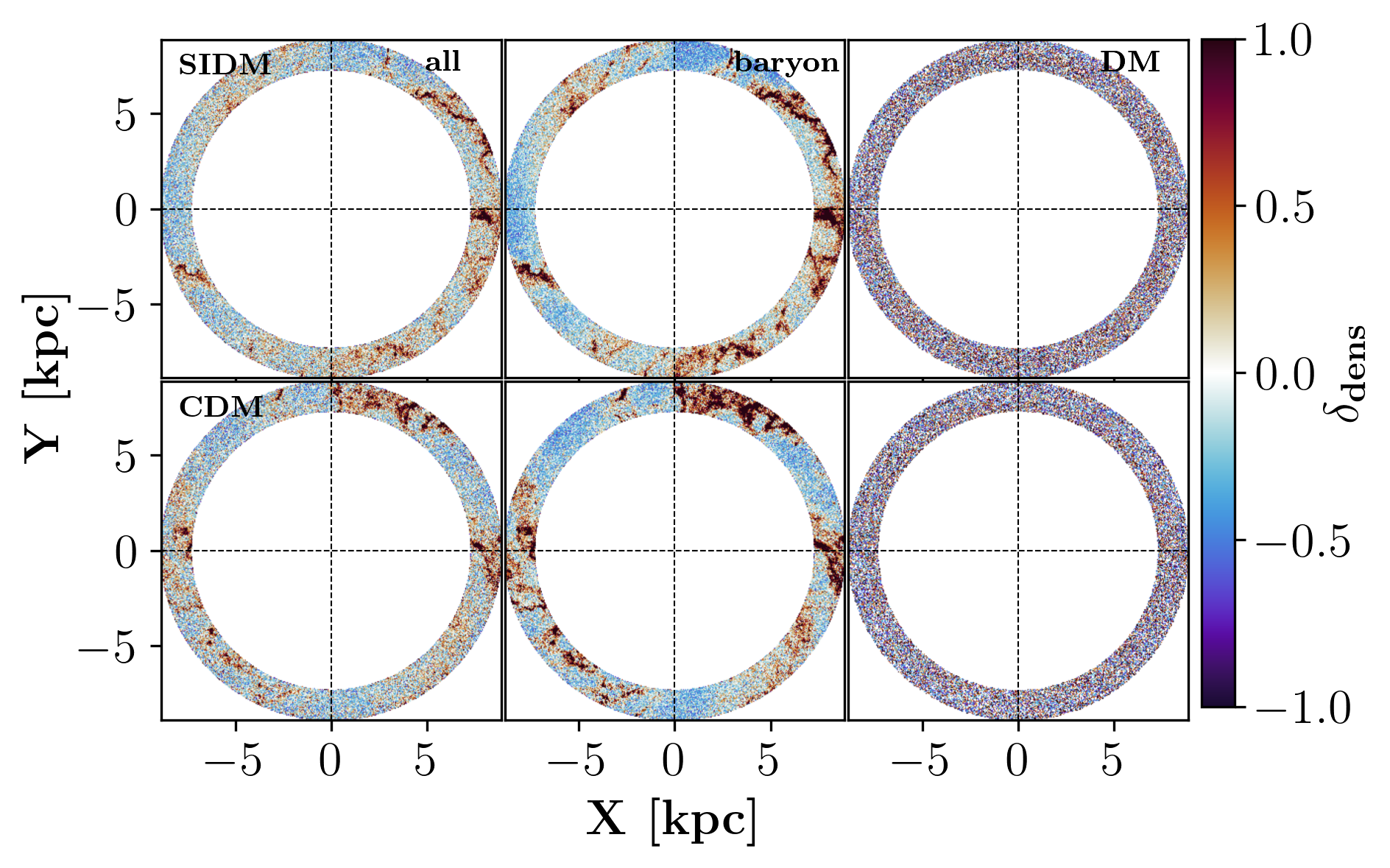}%
\caption{Relative variation in surface density ($\delta_\textrm{dens}$) in the XY plane around the Solar Neighborhood within $\textrm{R}_{\odot} = 8.1 \pm 0.6$ kpc and $|Z| < 5$ kpc for m12f halos; i. SIDM ({top}), and ii. CDM ({bottom}) computed for particles species (all--\textit{left}, baryons--\textit{middle}, DM--\textit{right}). Most variation with overdensities (red) and underdensities (blue) are in the baryonic component, while DM densities are relatively uniform and smooth.}\label{fig:m12f_clumps}
\end{figure*}

\setlength{\tabcolsep}{3.7pt}
\begin{table*}[ht]
\footnotesize
\caption{Properties of simulated galaxies and their azimuthal variation at the Solar Circle, $\textrm{R}_{\odot} = 8.1 \pm 0.6$ kpc, comparing CDM and SIDM models with the MW at present day. $\textrm{M}_\star$: Stellar mass within 10\% of the radius where the cumulative density is 200 times the critical density of the Universe. $\textrm{r}_{1/2}$: 3D stellar half-mass radius.}
\begin{center}   
\begin{tabular}{|c||c|cc|cc|cccc|ccc|}
\multicolumn{1}{l}{} & \multicolumn{1}{l}{} & \multicolumn{2}{c}{$\textrm{halo property}$} & \multicolumn{2}{c}{$\textrm{surface density}^\textrm{c}$ {[}$\textrm{M}_{\odot}/\textrm{pc}^{-2}${]}} & \multicolumn{4}{c}{$\textrm{volume density}^\textrm{d}$ {[}$10^{-3} \textrm{M}_{\odot}/\textrm{pc}^{-3}${]}} & \multicolumn{3}{c}{scale height {[}pc{]}} \\ \hline
\textbf{Galaxy} & \textbf{Physics} & $\textrm{M}_\star^\textrm{b} [\textrm{M}_{\odot}]$ & $\textrm{r}_{1/2}^\textrm{b} [\textrm{kpc}]$ & $\textrm{total}$ & $\textrm{baryonic}$ & total & stars & gas & DM & \multicolumn{1}{l}{\begin{tabular}[c]{@{}l@{}}stars\\ thin\end{tabular}} & \multicolumn{1}{l}{\begin{tabular}[c]{@{}l@{}}stars\\ thick\end{tabular}} & \multicolumn{1}{l|}{\begin{tabular}[c]{@{}l@{}}$\textrm{cold}^*$ \\ gas\end{tabular}} \\ \hline \hline

$\textrm{MW}^\textrm{a}$ & \textbf{?}  & ${5^{+0.4}_{-0.5} \times 10^{10}}$ & ${4.2^{+1}_{-1}}$ & 70$^{+5}_{-5}$ & 43.8$^{+3.4}_{-3.4}$ & 99.5$^{+9.5}_{-9.5}$ & 42.4$^{+3}_{-3}$ & 45.9$^{+8}_{-8}$ & 10.5$^{+2.5}_{-2.5}$ & 300$^{+60}_{-60}$ & 900$^{+180}_{-180}$ & 150 \\ \hline
\multirow{2}{*}{m12i} & CDM & $3.4 \times 10^{10}$ & 4.3 & 44.7$^{+24}_{-5}$ & 25.3$^{+24}_{-4}$ & 23.1$^{+15}_{-11}$ & 7.8$^{+6.3}_{-4.5}$ & 6.6$^{+8.2}_{-5.8}$ & 8.7$^{+0.2}_{-0.2}$ & 566.4 & $2000^\dag$ & 321.5 \\ 
& SIDM & $5.2 \times 10^{10}$ & 3.8 & 55.9$^{+24}_{-17}$ & 36.8$^{+23}_{-17}$ & 30.3$^{+18}_{-15}$ & 11.2$^{+5.2}_{-4.5}$ & 10.4$^{+13}_{-10}$ & 9.1$^{+0.1}_{-0.1}$ & 488 & 1503.4 & 359.1 \\ \hline
\multirow{2}{*}{m12f} & CDM & $5.8 \times 10^{10}$ & 3.6 & 61.1$^{+20}_{-13}$ & 39.2$^{+19}_{-11}$ & 33.5$^{+19}_{-11}$ & 12.9$^{+4.3}_{-3.1}$ & 10.4$^{+15.2}_{-7.6}$ & 10.1$^{+0.3}_{-0.3}$ & 569.9 & $2000^\dag$ & 363.4 \\
& SIDM & $6.7 \times 10^{10}$ & 4.7 & 67.9$^{+20}_{-16}$ & 43.4$^{+20}_{-16}$ & 35.7$^{+22}_{-10}$ & 16.7$^{+5.5}_{-3.2}$ & 7.6$^{+16}_{-6.4}$ & 11.4$^{+0.3}_{-0.1}$ & 437.6 & 1078.6 & 322.8 \\ \hline
\multirow{2}{*}{m12m} & CDM & $5.1 \times 10^{10}$ & 8.2 & 63.3$^{+19}_{-15}$ & 39.6$^{+18}_{-15}$ & 35$^{+20}_{-10}$ & 16$^{+5.7}_{-2.8}$ & 8.3$^{+13.2}_{-7.3}$ & 10.6$^{+0.2}_{-0.2}$ & 153.3 & 682 & 297.6 \\
& SIDM & $6.9 \times 10^{10}$ & 8.3 & 87.9$^{30}_{-26}$ & 57$^{+29}_{-25}$ & 49.5$^{+28}_{-13}$ & 25$^{+8.6}_{-5.4}$ & 10.7$^{+19}_{-7.4}$ & 13.9$^{+0.3}_{-0.3}$ & 218.4 & 742.9 & 317.2 \\ \hline                                    
\end{tabular}
\end{center}
\tablecomments{\label{tab:gal_prop} $^\textrm{a}$ values from \cite{mckee2015stars} unless noted, consistent within one standard deviation to values in \cite{bland2016galaxy}. $^\textrm{d}$ values from \cite{cautun2020milky}, consistent within one standard deviation to values in \cite{bland2016galaxy} and \texttt{MWPotential2022} from \texttt{GALA}  \citet{price2017gala}. $^\textrm{c}$ Materials within $|Z| \leq 1.1$ kpc. $^\textrm{d}$ Materials within $|Z| \leq 0.2$ kpc with values from \cite{Schutz2018,pablo2020}, the MW values are consistent within one standard deviation to values in \cite{mckee2015stars, bland2016galaxy}. $^*$Gas Temperature, $T \leq 100$ K. $^\dag$ Modeled better with single sech profile. Volume densities remain consistent with changing the width of the Solar Circle and $Z$ range. Please note that our halo values differ from those in \citet{sand2020_prop} because our CDM simulations, as noted in Sec.~\ref{sec:sims}, are different from the fiducial FIRE-2 suite. The error bars in the simulation show the azimuthal variation in low and high density regions.}
\end{table*}

Given the cosmological nature of these simulations in an arbitrary comoving box, we establish the galactocentric coordinates for each galaxy through a two-step process. First, we use the ``shrinking spheres'' method \citep{power2003inner} to determine the center position of each galaxy. We rotate the system to align the total angular momentum of the young stars (age $\leq 1$ Gyr) along the $Z$ direction, which orients the Galactic disk in the XY plane for each simulation.\footnote{We also tested other methods to establish the disk plane, such as measuring the spatial midplane, and found that our results remain consistent regardless of the method used.} 

We establish the Solar Circle with a fixed cylindrical radius of $\textrm{R}_{\odot} = 8.1 \pm 0.6$ kpc \citep{Gravity2018}. The rotation curves of these simulations are roughly flat at 8.1 kpc, with values close to $220 \pm 20$ km s$^{-1}$, which roughly match that of the MW. Our choice is strictly motivated from the measured Solar position, in Sec.~\ref{sec:dens_DM_radius}, we discuss our results for the DM density as a function of changing cylindrical radius from the center of each halo and show that the DM densities at $\textrm{R}_{\odot} = 8.1 \pm 0.6$ kpc are consistent with the MW values, within 1$\sigma$ (see Table~\ref{tab:gal_prop}). Although one could scale the cylindrical radius of each halo based on quantities like the exponential scale length of the Galactic disk or the stellar half-mass radius, the galaxies in this paper are within a factor of ~2 in stellar mass (see Table~\ref{tab:gal_prop}). Thus, scaling to a specific scale radius is less critical compared to cases where galaxies have widely different masses. The decision to scale depends on the properties being probed and whether there is a valid reason to expect them to scale with the scale radius at a fixed mass. For example, \citet{bellardini20213d} found that measuring FIRE-2 galaxies in physical units rather than scaling to a specific radius resulted in less scatter and more self-similarity in their analysis of metallicity radial gradients. Since, our results depend on the density at a fixed radius, which does not necessarily scale with the scale radius at a fixed mass, the meaningful ``scaling'' in this context is to match the local densities rather than adjusting based on a scale radius. We find that using a fixed radius from the center of each halo offers a more consistent and reliable comparison for our analysis. 

We select a cylindrical region with a fixed Galactocentric cylindrical radius of $\textrm{R}_{\odot} = 8.1 \pm 0.6$ kpc and a height of $|Z| \leq 5$ kpc, centered around the Solar Circle. This cylindrical region is then divided into small bins\footnote{This value is $\geq 10$ times larger than the star and gas softening parameters used in the simulations and about 2 times greater than the DM softening.} of 50 kpc in the X and Y direction, covering the $|Z| \leq 5$ kpc. Within each of these bins, we independently compute the surface density ($\Sigma_{X,Y}$) for baryons (stars and gas), DM, and for all (baryons and DM) the species combined together. Following this, we introduce a metric, $\delta_\textrm{dens}$, to quantify the relative surface density variation:
\begin{equation}
    \delta_\textrm{dens} = \frac{\Sigma_{X,Y} - \overline{\Sigma_{R}}}{\overline{\Sigma_{R}}}
\end{equation}
where $\overline{\Sigma_{R}}$ is the median surface density at $\mathrm{R = {(X^2+Y^2)}^{1/2}}$. 

Fig.~\ref{fig:m12f_clumps} illustrates relative surface density variation ($\delta_\textrm{dens}$) for m12f SIDM (top panel) and CDM (bottom panel) halos at the present day, categorized by species: all (left), baryons (middle), and DM (right). Major density variations are predominantly within the baryonic component, where overdense regions (red) can exhibit densities approximately twice as high ($\delta_\textrm{dens} \sim 1$). Conversely, the DM component showcases a uniform distribution, characterized by $|\delta_\textrm{dens}| < 0.05$. This uniformity in DM serves to counterbalance the fluctuations observed in baryonic matter. However, when considering all species combined, notable relative density variations persist within the Solar Neighborhood. Similar patterns are observed in m12i and m12m simulations. Notably, no discernible systematic trend in densities is observed between CDM and SIDM models. Furthermore, majority of the high density regions exhibit elevated gas density and are actively forming new stars. We note that $\delta_\textrm{dens}$ is roughly independent on the choice of $Z$ range.

Next, we categorize the bins into three density regions -- low, median, and high -- based on their surface density  in quartile ranges $\leq25^\textrm{th}$ percentile, $25^\textrm{th}-75^\textrm{th}$ percentile, and $\geq75^\textrm{th}$ percentile respectively. The three density regimes effectively provide upper and lower limits on the Galactic acceleration profiles.
Table~\ref{tab:gal_prop} summarizes the galaxy-wide properties along with average variations in the low, median, and high density regions around the Solar Circle for all of the simulations.\footnote{Varying $|Z|$ ranges to match the MW estimates.} While, the median baryonic volume density in the Solar Circle across all halos is generally much lower ($5-8\sigma$) compared to the MW, the total surface density across the halos is relatively close to that of the MW (within $2-3\sigma$), except for m12i CDM that is about $6\sigma$ away. Also, \citet{mccluskey2024disc} showed that the MW forms an unusually dynamically cold disk which could explain higher average density, contrasting with other observed galaxies of similar mass ($\sim 10^{12}$ \Msol{}) through analysis of the \textit{Latte} disks. Additionally, our estimates for the MW are solely based on the local volume around the Solar Neighborhood. In terms of matching local properties, the m12m SIDM halo is the most similar to the MW, exhibiting comparable thin and thick disk, and cold gas scale heights. 
We observe that both the total volume and surface density between regions within a same halo can vary by a factor 2. For example, in the low, median, and high density regions, the volume density\footnote{for $|Z| \leq 0.2$ kpc to match the volume density calculation in the MW} for m12m SIDM is approximately (36, 50, 79) $\cdot 10^{-3} \textrm{M}_{\odot}/\textrm{pc}^{-3}$. Only $\sim6\%$ of our selected bins exhibit total volume densities within $1\sigma$ of the MW value of $99 \pm 9.5 \ \cdot 10^{-3}\textrm{M}_{\odot}/\textrm{pc}^{-3}$ and 10\% bins have total volume density values higher than the MW. The majority of the variations arise from the baryonic component (stars and gas) and the DM component is evenly distributed azimuthally (lower errors in Table~\ref{tab:gal_prop}, see Fig.~\ref{fig:m12f_clumps}.).
 
Comparing DM models, the SIDM halos exhibit both higher stellar and DM volume and surface densities in the Solar Neighborhood than their CDM counterparts. This is primarily due to SIDM particles' capacity to exchange energy and momentum in addition to gravity, making them more responsive and sensitive to the baryonic potential \citep{sameie2018impact,elbert2018testable,despali2019interplay,robles2019milky, santos2020baryonic}. The higher DM density facilitates the entrapment of additional gas, forming new stars and establishing a feedback loop, where DM particles can respond faster to the rising stellar density \citep{despali2019interplay, sameie2021central}. While at about 8.1 kpc from the Galactic center, we anticipate only about one DM scattering event per particle per Hubble time, these rates are higher in the inner regions of the galaxy \citep{vargya2022shapes},  which leads to  more pronounced differences in DM density between CDM and SIDM in the inner regions (see Fig.~\ref{fig:dens_dm_rad} in Sec.~\ref{sec:dens_DM_radius}). Consequently, SIDM halos exhibit higher density in DM and stellar distribution within the baryon-dominated potential, resulting in increased oblateness in the inner regions and around the Solar Circle.  

In Sec.~\ref{sec:dens_DM_radius} (see Fig.~\ref{fig:dens_dm_rad}), we show the azimuthally average DM density in different 2D cylindrical radii (between 4 to 12 kpc), showing that the trend of denser SIDM holds across a variety of radii. The differences between the CDM and SIDM are more pronounced in the inner regions, and considering that the MW has a thinner disk compared to the simulated galaxies, the expected effect of SIDM versus CDM as a function of vertical distance from the midplane ($Z$) in the MW should be even more significant.

\subsection{Comparison with local potential models} \label{sec:oblt}

\setlength{\tabcolsep}{4pt}
\begin{table*}
\caption{Best fit potential model and oblateness/shape parameters.}
\begin{center}
\begin{tabular}{|c||c|cccc|cc|c|}
\hline
\textbf{Galaxy} & {\textbf{Physics}} & \begin{tabular}[c]{@{}l@{}}$\mathrm{M_{MN}}$\\ {[}$10^{10} \textrm{M}_{\odot}${]}\end{tabular} & \begin{tabular}[c]{@{}l@{}}a\\ {[}kpc{]}\end{tabular} & \begin{tabular}[c]{@{}l@{}}b\\ {[}kpc{]}\end{tabular} & \multicolumn{1}{l|}{$\mathrm{\log_{10}(-\gamma_{MN}/Gyr^{-2})}$} & \multicolumn{1}{l}{$\mathrm{\log_{10}(\alpha_{1}/Gyr^{-2})}$} & \multicolumn{1}{l|}{$\mathrm{\log_{10}(-\gamma_{LO}/Gyr^{-2})}$} & \multicolumn{1}{l|}{$\frac{c}{a}\Big\rfloor_{\textrm{R}_{\odot}}$} \\ \hline \hline

$\textrm{MW}^\textrm{a}$ & \textbf{?} & 10 & 6.5 & 0.26 & 3.94 & $3.54 \pm 0.16$ & $3.3 \pm 0.9$ & -- \\ \hline
\multirow{2}{*}{m12i} & CDM & $1.95 \pm 0.04$ & $3.91 \pm 0.31$ & $0.80 \pm 0.12$ & $2.83 \pm 0.02$ & $3.00 \pm 0.00$ & $2.87 \pm 0.01$ & 0.46 \\
& SIDM & $3.11 \pm 0.07$ & $3.17 \pm 0.34$ & $0.85 \pm 0.23$ & $3.01 \pm 0.03$  & $3.14 \pm 0.01$ & $2.99 \pm 0.01$ & 0.43 \\ \hline

\multirow{2}{*}{m12f} & CDM & $3.54 \pm 0.13$ & $3.30 \pm 0.26$ & $0.82 \pm 0.33$ & $3.08 \pm 0.05$ & $3.17 \pm 0.02$ & $3.15 \pm 0.00$ & 0.40\\
& SIDM & $4.20 \pm 0.05$ & $3.50 \pm 0.18$ & $0.79 \pm 0.05$ & $3.17 \pm 0.02$ & $3.20 \pm 0.00$ & $3.16 \pm 0.01$ & 0.38 \\ \hline

\multirow{2}{*}{m12m} & CDM & $3.49 \pm 0.24$ & $6.31 \pm 0.44$ & $0.78 \pm 0.04$ & $2.97 \pm 0.04$ & $3.17 \pm 0.01$ & $2.92 \pm 0.00$ & 0.30 \\
& SIDM & $5.30 \pm 0.09$ & $6.81 \pm 0.20$ & $0.78 \pm 0.04$ & $3.11 \pm 0.05$  & $3.28 \pm 0.01$ & $2.95 \pm 0.01$ & 0.27 \\ \hline        
\end{tabular}
\end{center}
\tablecomments{Miyamoto-Nagai profile parameters for the stellar disk for stars within $3 \textrm{kpc} \leq \textrm{R} \leq 15 \textrm{kpc}$ and $|Z| \leq 1.1$ kpc. LO potential profile parameters near the Solar Circle, $\textrm{R}_{\odot} = 8.1$ kpc for stars within $|Z| \leq 1.1$ kpc, for the simulated galaxies with different DM models. $\frac{c}{a}\Big\rfloor_{\textrm{R}_{\odot}}$ is the minimum to major axis ratio computed using an ellipsoid fit to all matter particles at $\textrm{R}_{\odot}$. \\ $^\mathrm{a}$ values from \cite{chakrabarti2021measurement, donlon2024galactic}.}\label{tab:gal_potential}
\end{table*}

Ideally, one would integrate the acceleration field (from e.g., pulsar timing or other acceleration measurements) to infer the potential directly. However, in practice, the scarcity of data points (e.g., 20 pulsars within 2 kpc of the Sun \citep{donlon2024galactic}) currently limits our ability to perform such integration. Consequently, in the MW, observers use static parameterize models to fit the available data accurately. Here, we present our simulations with similar parameters through potential model fitting on stars in order for a direct comparison. We consider two axisymmetric potentials previously used in \cite{chakrabarti2021measurement}. Specifically, a low-order (LO) expansion $\alpha \gamma$ potential \citep[eq.~\ref{eq:LO_pot},][]{chakrabarti2021measurement} near the position of the sun (i.e adopted Solar Circle in our case), and a single Miyamoto-Nagai (MN) potential (eq.~\ref{eq:MN_pot}) for the Galactic disk. 

The LO potential profile is defined as
\begin{equation} \label{eq:LO_pot}
\Phi(R, Z)=V_{\mathrm{LSR}}^2 \log \left(R / R_{\odot}\right)+\log \left(R / R_{\odot}\right) \gamma_\mathrm{LO} Z^2+\frac{1}{2} \alpha_1 Z^2
\end{equation}

where $V_{\mathrm{LSR}}$ is the local standard of rest velocity, $\gamma_\mathrm{LO}$ determines the shape of the potential affecting the vertical accelerations, and $\alpha_1$ represents the squared angular frequency of oscillations in the vertical density profile. A higher value of  $\log{-\gamma_\mathrm{LO}}$ value indicates a flatter potential profile in the Solar Neighborhood. 

For stars located within $\textrm{R}_{\odot} = 8.1 \pm 0.6$ kpc and at heights $|Z| \leq 1.1$ kpc, localized around the midplane, we fit the LO potential profile paramters $V_{\mathrm{LSR}}$, $\gamma_\mathrm{LO}$, and $\alpha_1$ using linear regressions on the potential values stored as a property for the particles in the simulation. To estimate the uncertainties on these parameters, we employ a Markov Chain Monte Carlo (MCMC) bootstrap approach, using 1000 samples. This involves repeatedly resampling the data with replacement and fitting the model to each resampled dataset. 

To model the gravitational effect of the simulated galactic disk for stars within $3 \textrm{kpc} \leq \textrm{R} \leq 15 \textrm{kpc}$ and at heights $|Z| \leq 1.1$ kpc, we use a single MN profile for the limited vertical range around the midplane, given by

\begin{equation} \label{eq:MN_pot}
\Phi_\mathrm{MN}(R,Z)=-\frac{G M_\mathrm{MN}}{\sqrt{R^{2}+(a+\sqrt{Z^{2}+b^{2}})^{2}}} .
\end{equation}

We fit the 2D density of the simulation with the density functional form of the MN profile using the scale mass ($M_\mathrm{MN}$), scale radius ($a$), and scale height ($b$) for the Galactic disk. The fitting is performed using a least squares optimization method, specifically the Nelder-Mead method \citep{neldermead1998convergence}, to find the best fit parameters and estimate the uncertainties on these parameters using MCMC bootstrap using 1000 samples. \citet{arora2022stability} showed that such analytic potential models can reconstruct the rotation curve of the FIRE simulated galaxies to about $\pm2\%$ accuracy in the disk region. 

The oblateness parameter for the MN disk is computed by expanding eq.~\ref{eq:MN_pot} to first order in $R$ and second order in $Z$ around ($R_{\odot}, 0$):  

\begin{equation}
    \gamma_{\mathrm{MN}}=-\frac{G M_{\mathrm{MN}}}{b} \frac{a+b}{\left(R_{\odot}^2+(a+b)^2\right)^{5 / 2}} \frac{3 R_{\odot}^2}{2} .
\end{equation}

Table~\ref{tab:gal_potential} shows the best fit parameters and oblateness ($\gamma$) obtained for the simulations using the LO and MN potentials along with the best fit parameters for the MW from \cite{donlon2024galactic}. The scale radius ($a$), and scale height ($b$) from the MN profile fits for all three halos are within 1$\sigma$ of each other when comparing the CDM and SIDM models. It is noteworthy that the $\gamma$ values derived from the LO potential ($\gamma_\mathrm{LO}$) and the MN disk ($\gamma_\mathrm{MN}$) are similar in all simulations, except for the m12m SIDM case. This suggests that the shape of the local potential at the Solar Circle is primary influenced by the stellar disk. 

The values obtained in the simulations are within $0.5\sigma$ of the MW best fit parameters from \citep{donlon2024galactic}. Our simulated SIDM halos match the observed MW values better than the CDM counterparts, particularly with regard to the shape and oblateness of DM halos. Specifically, the SIDM halos at the Solar Circle demonstrate lower minor to major axis ratios (by 0.02-0.03); in other words, they are more oblate than the CDM halos. This happens because SIDM magnifies the variations in SFR that enhances the concentration of both baryons and DM (see Table.~\ref{tab:gal_prop}). 


\section{Vertical acceleration profiles} \label{sec:gal_acc}
\begin{figure}
\includegraphics[width=\linewidth]{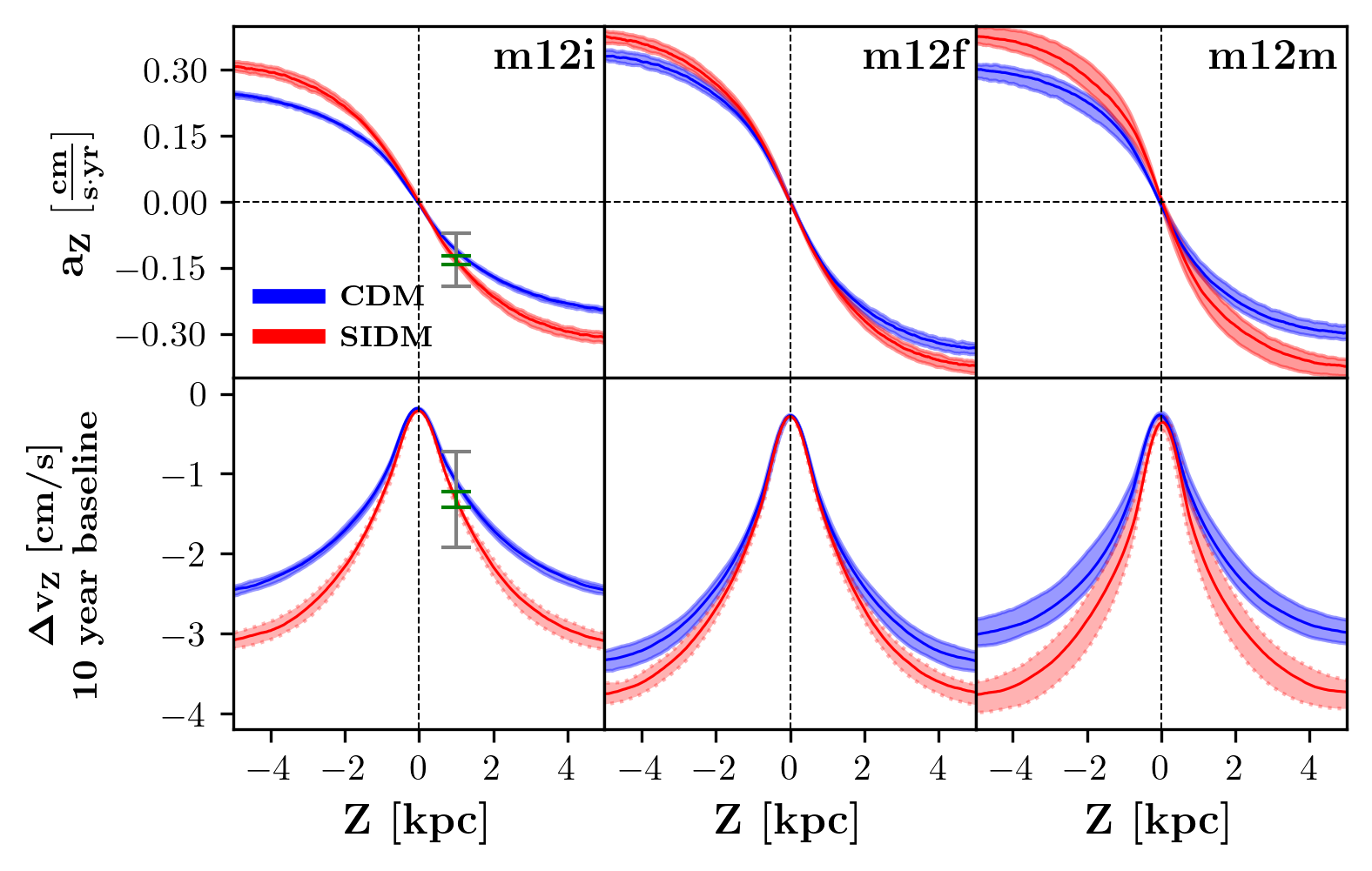}%
\caption{Vertical acceleration (top row) and change in the vertical velocity over a 10-year baseline (bottom row) as a function of vertical height $Z$ from the Solar Circle for CDM (blue) and SIDM (red) models in the three simulations (columns). The corresponding shaded regions show the upper and lower limits on the profiles, as computed from high and low density bins respectively. The error bars in the first column show the current (with a 10 year baseline) measurement uncertainties for pulsar accelerations up to $Z=\pm1$ kpc (gray), and expected uncertainties with a 20 year baseline (green) from \citet{donlon2024galactic} by scaling pulsar timing
precision by a power of 5/3 \citep{Lorimer_Kramer2012} centered on the m12i SIDM value at $Z=1$ kpc for reference. It is evident that both the vertical acceleration profile and the change in vertical velocity have consistently larger values in the SIDM simulations compared to their CDM counterparts.}\label{fig:acc_vel}
\end{figure}

We calculate the vertical accelerations above and below the Galactic midplane using a binned method ($\Delta Z = 0.05$ kpc) to present smooth trends as a function of vertical distance ($Z$) from the midplane.
We slice the low, median, and high density regions as defined in Sec.~\ref{sec:solar_neigh} into 50 kpc vertical distance $Z$ bins\footnote{This value is $\geq 10$ times greater than the softening parameters used in the simulations.}. Next, we average the gravitational accelerations computed for stars and DM particles in each region within each $Z$ bin. With this approach, we obtain a smooth and accurate representation of the median acceleration profile; the high density bin provide an upper limit, and the low density bin provide a lower limit for each profile. 

Fig.~\ref{fig:acc_vel} plots the vertical acceleration $\mathrm{a_Z}$ (top row) and change in the vertical velocity over a 10-year baseline ($\mathrm{\Delta v_{Z} \equiv a_{Z}\Delta t}$) (bottom row) at present day for the m12i (left column), m12f (middle column), and m12m (right column) in CDM (blue) and SIDM (red) model. The shaded regions represent the upper and lower limits on the profiles derived from the high and low density regions respectively. SIDM halos have consistently larger magnitude of their accelerations than CDM, thanks to their higher surface densities, especially at larger distances from the mid-plane. The error bars in the first column represent current average measurement uncertainties for pulsar accelerations up to $Z=\pm1$ kpc (gray error bars) centered on the m12i SIDM value at $Z=1$ kpc for reference, derived from pulsar timing data over a 10-year baseline, and the expected uncertainties for a 20-year baseline (green error bars) as reported in \citet{donlon2024galactic} scaling pulsar timing
precision by a power of 5/3 \citep{Lorimer_Kramer2012}. The estimated future uncertainties are expected to distinguish between the profile differences noted between the CDM and SIDM models. We note that the degree of difference in $\mathrm{\Delta v_{Z}}$ between SIDM and CDM profiles varies with the simulation, with the smallest difference observed for m12f (one with a massive merger about 3 Gyr ago before the present day) and the largest difference for m12m (no mergers in the last 10 Gyr). This observation suggests that mergers may have a role in removing or mitigating differences between SIDM and CDM profiles. We discuss this more in Sec.~\ref{sec:m12f_acc_time}.

\subsection{Median vertical acceleration gradient} \label{sec:da_dz_mean}

\begin{figure*}
\includegraphics[width=\textwidth,]{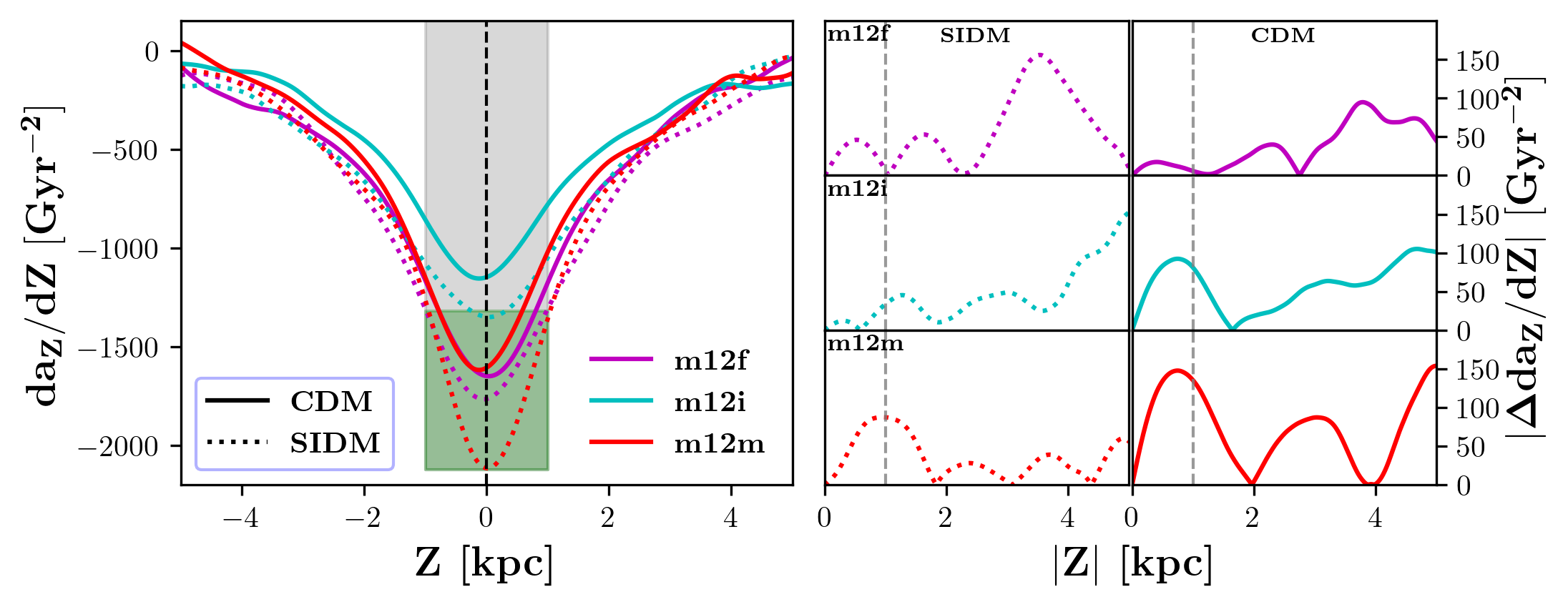}%
\caption{Vertical acceleration gradient ($\mathrm{\frac{da_Z}{dZ} (Z)}$) (left) and the asymmetric difference between $+Z$ and $-Z$ (right) with respect to vertical height $Z$ for CDM (solid) and SIDM (dotted) halos. SIDM halos have consistently steeper profiles compared to the CDM halos. Notable asymmetries are observed in m12f and m12i for $Z \geq 1$ kpc (dashed line in second panel). The MW's measured $\mathrm{\frac{da_Z}{dZ}}$ at midplane is approximately -3200 $\textrm{Gyr}^{-2}$. Gray shading represents the current measurement uncertainty for pulsar studies in $\mathrm{\frac{da_Z}{dZ}}$ \citep{chakrabarti2021measurement, donlon2024galactic}, and the green region indicates the expected precision increase, scaling pulsar timing precision by a power of 5/3 over a 20 year baseline \citep{Lorimer_Kramer2012}. Both the shaded regions are scaled to match the minimum value in m12m SIDM. Fig.~\ref{fig:dens_dm_rad} in Sec.~\ref{sec:dens_DM_radius} plots the $\mathrm{da_Z/dZ (Z=0)}$ at midplane ($Z=0$) as a function of 2D cylindrical radius. SIDM halos show consistently larger value of $\mathrm{\frac{da_Z}{dZ} (Z=0)}$ across a range of radii. The differences between CDM and SIDM are more pronounced closer to the galactic center.}\label{fig:da_dz}
\end{figure*}

Fig.~\ref{fig:da_dz} depicts the vertical acceleration gradient ($\mathrm{\frac{da_Z}{dZ}}$) (left) and the asymmetric difference around the midplane (right) as a function of vertical height $Z$ for simulations (color-coded) in CDM (solid line) and SIDM (dotted line) models. The gradients are computed using total-variation regularization differentiation algorithm which avoids the noise amplification in finite-difference methods for noisy data \citep{chartrand2011numerical}. The asymmetric difference is given by 

\begin{equation} \label{eq:asymm_diff}
    \Bigg |\Delta\mathrm{\frac{da_Z}{dZ}} \Bigg | \equiv \Bigg | \mathrm{\frac{da_Z}{dZ} \Big \rfloor_{+Z} - \frac{da_Z}{dZ} \Big \rfloor_{-Z}} \Bigg |
\end{equation}

The gray shaded region indicates the current measurement uncertainty in $\mathrm{\frac{da_Z}{dZ}}$ from \cite{donlon2024galactic}, while the green shaded region represents the anticipated measurement uncertainty with a 20-year baseline \citep{Lorimer_Kramer2012}. Note, the majority of the azimuthal variation stems from the \emph{observable} baryonic component and the azimuthal distribution of DM is effectively constant in these simulations. In fact, variations arising from the baryonic component can be mitigated with precise measurements, as such we only show the median profiles for $\mathrm{\frac{da_Z}{dZ}}$ in Fig.~\ref{fig:da_dz}, ~\ref{fig:m12f_time}, and ~\ref{fig:dens_dm_rad}.  

The $\mathrm{\frac{da_Z}{dZ}}$ at the midplane ($Z = 0$ kpc) in our simulations closely aligns with the MW's best-fit value of $-3200 \pm 2560$ Gyr$^{-2}$ \citep{donlon2024galactic}, albeit within $1\sigma$ error of the MW measurement. Discrepancies primarily arise from lower median total matter density in the Solar Neighborhood in the simulations relative to the measured MW values that is locally measured. \citet{chakrabarti2021measurement} reported a value of $-4900^{+1600}_{-2700}$ Gyr$^{-2}$ using data from 14 binary pulsars distributed over $\sim$ 1 kpc from the Sun, which is comparable to the more updated value given in \citet{donlon2024galactic}. This value corresponds to the square of the frequency of low-amplitude vertical oscillations (denoted $\alpha_{1}$ in \citet{chakrabarti2021measurement}), and is used to determine the mid-plane density, or the Oort limit.  This gives a value of the Oort limit of $0.08^{0.05}_{-0.02}~M_{\odot}/\rm pc^{3}$, which is close to but lower than kinematic estimates. Due to the limitation of data available at that time, the analysis in \citet{chakrabarti2021measurement} could not constrain global properties, like the mass.  
\citet{donlon2024galactic} find a value for the Oort limit of $0.062\pm0.017~M_{\odot}/\rm pc^{3}$, which is lower still relative to kinematic estimates.  
Using simplified density profiles, \citet{donlon2024galactic} estimates the galaxy mass within the Solar Circle to be nearly twice as high as the currently accepted value from \citet{bland2016galaxy}.  MW measurements may however be impacted by high uncertainties currently in binary pulsar data.
These measurements will continue to improve as more pulsar timing data become available.     

Notably, the depth of the $\mathrm{\frac{da_Z}{dZ} (Z)}$ as a function of vertical distance from the midplane varies significantly across simulations, with SIDM simulations consistently displaying deeper valleys by 10-30\% within 1 kpc of the midplane. The most substantial gradient and the most pronounced difference between SIDM and CDM profiles are observed in m12m because it is the most massive disk (interestingly, it also has the lowest Toomre Q value). In contrast, m12i CDM displays the shallowest valley close to the midplane, aligning with the volume densities of their respective systems. Interestingly, m12f, which experienced a recent merger, shows the least pronounced difference between CDM and SIDM profiles, while m12m, which had the earliest merger, exhibits the most significant difference between the DM profiles close to the midplane. In the outskirts ($|Z| \geq 2$ kpc), the difference between profiles is smaller. This observation suggests that the presence of recent major mergers such as Sag dSph and the LMC in a galaxy might mitigate the distinctive signatures of DM models on the vertical acceleration gradient profile, particularly close to the mid-plane. This could be due to multiple factors, such as an epoch of increased star formation triggered by the influx of gas from the merging satellites \citep{di2007star, hopkins2010mergers, pearson2019effect}, and/or heating of the galactic center due to the energy exchange with the merging satellite \citep[e.g.,][]{barnes1992dynamics}.

Given the uncertainties regarding the exact location of the Solar Circle within our simulation, Sec.~\ref{sec:dens_DM_radius}, Fig.\ref{fig:dens_dm_rad} shows $\mathrm{\frac{da_Z}{dZ}}$ as a function of 2D cylindrical radius spanning a range of 4--12 kpc. SIDM halos have consistently higher values of $\mathrm{\frac{da_Z}{dZ} (Z=0)}$ across all radii. The difference between SIDM and CDM model is most pronounced near the galactic center. Notably, while m12m exhibits the most significant difference, m12f displays relatively minor disparities, aligning with the trends observed in Fig.\ref{fig:da_dz}. These highlight the consistency of our results across a broad range of radii.

We also note a significant asymmetry around the mid-plane in $\mathrm{\frac{da_Z}{dZ}}$ (right panels in Fig.~\ref{fig:da_dz}), particularly pronounced for m12f, which underwent recent massive mergers, and also for m12i. In contrast, m12m, which had no recent mergers in the CDM case, exhibits a less pronounced asymmetry. For SIDM simulations, a similar trend is observed between m12f and m12i, with noticeable asymmetries in m12m as well. 

These asymmetries are influenced by various factors, including the orbits of merging satellites, which can excite long-standing nodes in the Galactic disk \citep{weinberg1998dynamics}, leading to higher variation in azimuthal direction. As argued by \cite{Chakrabartietal2019, chakrabarti2020toward}, recent mergers leave a discernible imprint on the vertical acceleration profiles persisting long after the complete tidal stripping of satellites. 
These signatures become more pronounced at greater distances from the mid-plane, highlighting the enduring impact of recent merger events.  We observe large vertical asymmetries especially in the gas disk in the outer part of the MW \citep{Levineetal2006} that may arise from an interaction with a massive dark matter sub-halo \citep{Chakrabarti_Blitz2009}. The \emph{global} asymmetries in the baryonic component, may allow us, together with the observed asymmetry in the total acceleration field, to more comprehensively model past interactions with dwarf galaxies.

\subsection{Temporal and spatial variation in the vertical acceleration gradient.} \label{sec:m12f_acc_time}

\begin{figure*}
\includegraphics[width=\textwidth,]{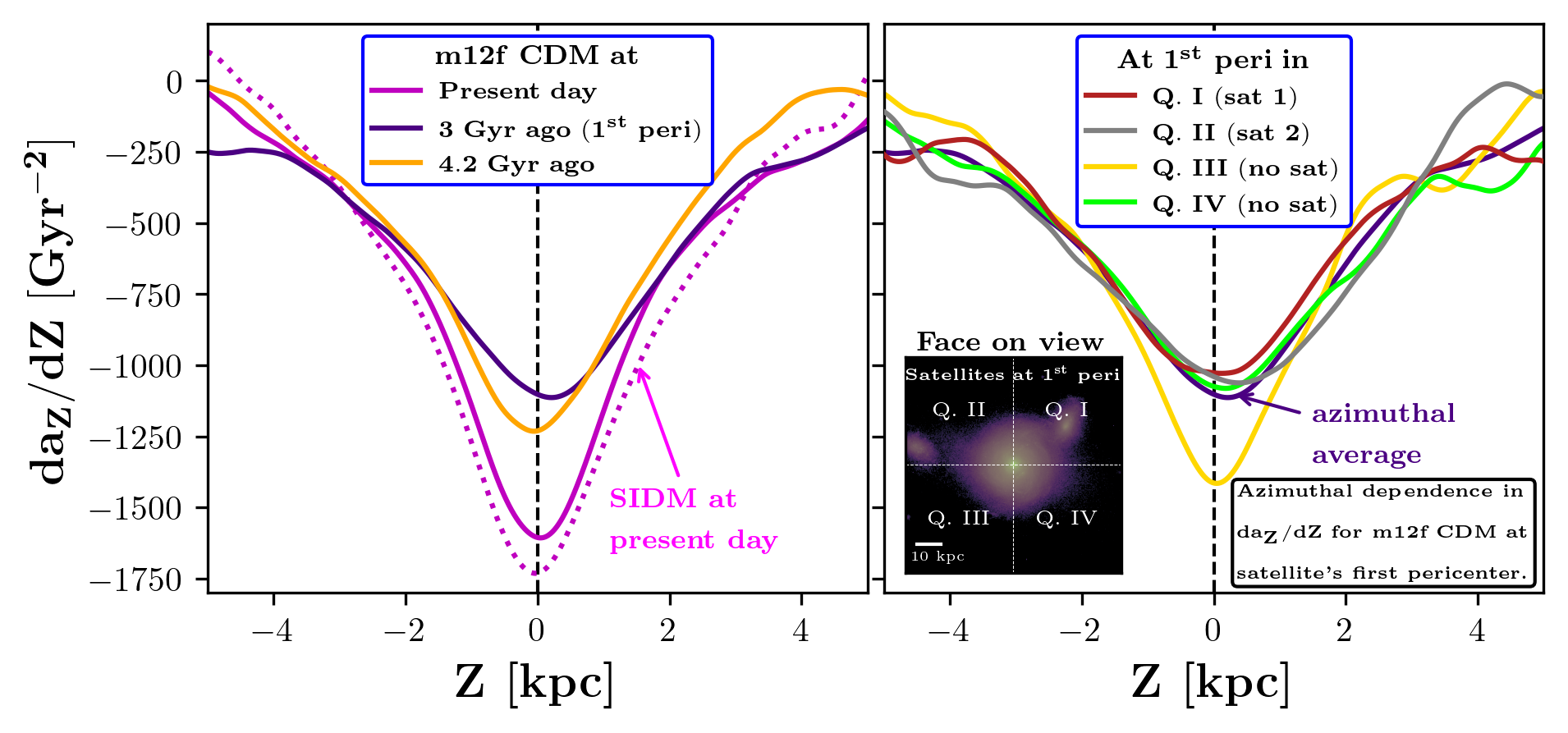}%
\caption{\textbf{Left:} Azimuthally averaged vertical acceleration gradient ($\mathrm{\frac{da_Z}{dZ}}$) as a function of vertical height $Z$ for the m12f CDM halo at different epochs: at present day (magenta color) for CDM (solid) and SIDM (dotted), at the time of the satellite's first pericentric passage (3 Gyr before the present day, indigo color), and 4.2 Gyr before the present day (orange color). Before the merger event, the profile is symmetric across the mid-plane ($Z=0$), but it becomes highly asymmetric during the merger and retains some asymmetry until the present day. \textbf{Right:} The vertical acceleration gradient ($\mathrm{\frac{da_Z}{dZ}}$) computed in different quadrants (marked in insets in the bottom left corner) for the m12f CDM halo at the time of the first pericentric passage. Q. I and II, which contain two satellites close to peri, exhibit the most asymmetry across the midplane. In contrast, Q. III and IV, which have no satellites, display relatively symmetric profiles. The inset in bottom left plots the disk plane for the m12f CDM at first pericenter and marks the relevant quadrants.} \label{fig:m12f_time}
\end{figure*}

In previous sections, we observed a consistent trend: azimuthally averaged vertical acceleration gradient profiles in SIDM simulations were systematically steeper, typically by 10-30\%, compared to their CDM counterparts. However, it remains whether temporal and azimuthal variations expected in $\mathrm{\frac{da_Z}{dZ} (Z)}$ are similar in level and scale as the DM model dependent differences because if this is true, it could imply that the variations we see in the simulations might not be due to actual differences in DM properties, but are actually transient effects due to mergers. In this section, we examine the spatial and temporal variations in the $\mathrm{\frac{da_Z}{dZ} (Z)}$ using the m12f CDM simulation as an example. This simulation experienced a major merger with two satellites with their first pericentric passage approximately 3 Gyr ago. With the total mass ratio\footnote{defined as the ratio of total mass of the main halo divided by the total mass of the satellite at the moment of first pericentric passage} of 15 and 18, when each satellite was at pericentric distance of about 25 kpc.   

The left panel of Fig.~\ref{fig:m12f_time} illustrates the azimuthally averaged $\mathrm{\frac{da_Z}{dZ} (Z)}$ profiles at various epochs: present day (magenta) both for CDM (solid) and SIDM (dotted), during the first pericentric passage (3 Gyr before the present day, indigo), and 4.2 Gyr before the present day (orange). These profiles display significant variations (10-30\%) in depth, influenced by the average total matter density. Approximately 1 Gyr before the pericentric passage, the profile is steeper due to enhanced star formation in the region around the Solar Circle and symmetric around the mid-plane. At the first pericentric passage, the profile is shallower and highly asymmetric. By the present day, the increasing baryon density has substantially steepened the acceleration profile, with the asymmetry persisting to some extent. These temporal variations in profile depth are comparable to the highest order variations seen in m12m CDM and SIDM profiles in Fig.~\ref{fig:da_dz} and higher than the variations observed in the the m12f CDM and SIDM profiles. 

The right panel of Fig.~\ref{fig:m12f_time} depicts $\mathrm{\frac{da_Z}{dZ} (Z)}$ computed in four azimuthal quadrants (marked in insets on the bottom left) during the first pericentric passage in m12f CDM. All quadrants exhibit consistent profile depths within a 10\% range, except for Q. III, which shows increased depth due to high-density gas prompting new star formation. The variation at large $|Z|$ stem from limited particle data away from the midplane leading to noisy gradients. Quadrants containing the two satellites (Q. I and II) display more significant asymmetry across the midplane, while the quadrants without any satellites (Q. III and IV) have relatively symmetric profiles, highlighting the impact of satellite orbits on gradient asymmetry.    

\subsubsection{Radial variation in the local DM density}\label{sec:dens_DM_radius}

\begin{figure*}
\includegraphics[width=\linewidth]{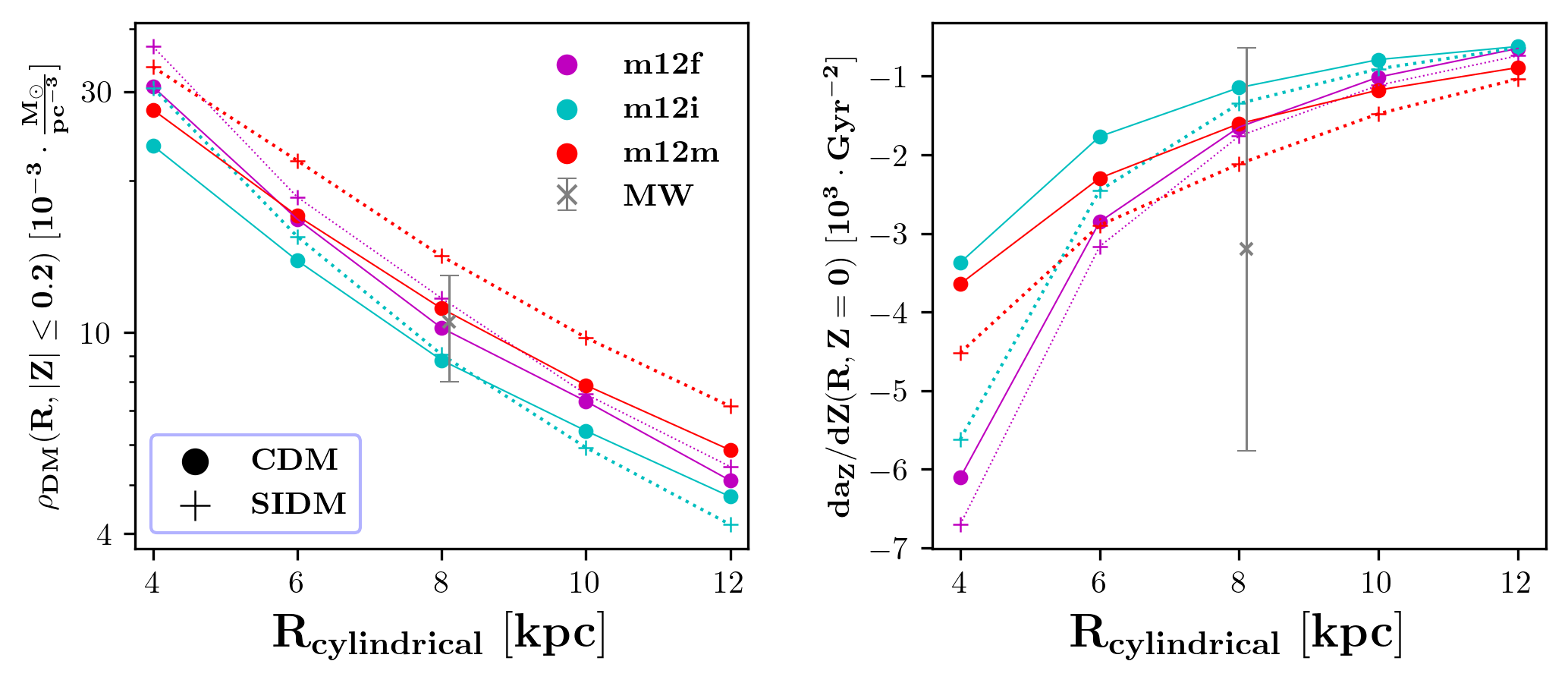}%
\caption{\textbf{Left:} Azimuthally averaged DM density ($\rho_\textrm{DM}$) within a vertical range of $|Z| \leq 0.2$ kpc as a function of 2D cylindrical radius ($R_\textrm{cylindrical}$) for: m12f (magenta), m12i (cyan), and m12m (red) for CDM (solid circles), and SIDM (plus) at present day. In general, SIDM halos have a higher DM density closer to the galactic center. Notably, m12m SIDM maintains consistently higher density compared to the CDM, whereas m12i halo, with the lowest stellar mass (see Table~\ref{tab:gal_prop}), shows a transition where SIDM becomes less dense (puffier) compared to CDM close to 8 kpc. \textbf{Right:} Vertical acceleration gradient ($\mathrm{\frac{da_Z}{dZ} (Z=0)}$) at the midplane ($Z=0$ kpc) as a function of cylindrical radius ($R_\textrm{cylindrical}$) for all the halos as in the left panel. The trends in $\mathrm{\frac{da_Z}{dZ} (Z=0)}$ are similar to those observed in DM density, with SIDM consistently exhibiting steeper gradients compared to CDM. The differences between SIDM and CDM are more pronounced closer to the center. Additionally, the current MW DM density \citep{bland2016galaxy} and $\mathrm{\frac{da_Z}{dZ} (Z=0)}$ \citep{donlon2024galactic} are shown with gray crosses with error bars.} \label{fig:dens_dm_rad}
\end{figure*}

Fig.\ref{fig:dens_dm_rad} plots the azimuthally averaged DM density\footnote{Similar to Table.\ref{tab:gal_prop}} ($\rho_\textrm{DM}$) calculated within a vertical range of $|Z| \leq 0.2$ kpc (left) and vertical acceleration gradient ($\mathrm{\frac{da_Z}{dZ} (Z=0)}$) at midplane ($Z=0$ kpc) as a function of 2D cylindrical radius ($R_\textrm{cylindrical}$) for the three simulations (color-coded) in both CDM (solid circles) and SIDM (plus) models. In the inner regions of the galaxies, SIDM models, influenced by the baryonic component, exhibit systematically higher DM density and steeper gradients at the mid-plane compared to their CDM counterparts for all radii. These differences decrease further away from the center. These pronounced differences are consistent with the higher scattering rates of DM particles close to the center \citep{vargya2022shapes}.  

In m12m, SIDM maintains more pronounced differences between densities and $\mathrm{\frac{da_Z}{dZ} (Z=0)}$ even at larger $R_\textrm{cylindrical}$, with no sign of convergence between CDM and SIDM up to 12 kpc. Conversely, in m12i, beyond 8 kpc, the SIDM profile becomes less dense than the CDM profile. This ``puffier'' profile is due to the absence of a strong baryonic potential in the outskirts, allowing DM particles to self-interact and form less dense, rounder profiles. Additionally, there are minor differences in the $\mathrm{\frac{da_Z}{dZ} (Z=0)}$ at larger $R_\textrm{cylindrical}$.Furthermore, m12f exhibits the least variation between CDM and SIDM at all radii in both density and $\mathrm{\frac{da_Z}{dZ} (Z=0)}$. However, $\mathrm{\frac{da_Z}{dZ} (Z=0)}$ remains systematically steeper in the SIDM model. 

The galatic disks in these simulated halos are thicker compared to the MW's disk \citep{wetzel2023public, mccluskey2024disc}. This implies that the effects of SIDM versus CDM could be even more pronounced in a thinner disk like that of the MW. 


\section{Discussion and conclusions} \label{Sec:disc}
Direct measurement of Galactic accelerations presents a promising avenue for testing DM, bypassing the need to solve the ``inverse problem'' \citep{binney2011galactic} inherent in traditional methods of inferring the mass distribution within galaxies from observed motions. By comparing observed acceleration profiles with predictions from simulations with different DM models, we can assess the compatibility of these models with observational data. However, testing SIDM in particular with this strategy offers some challenges due to the ability of SIDM to respond efficiently to changes in the baryon distribution. Galaxy formation can alter even the shape and density profile of a CDM halo; in SIDM, which can more easily alter its energy and angular momentum through interactions, the expectation is that the halo's shape and profile are more tightly correlated with the disk properties \citep{sameie2021central,vargya2022shapes}. However, this response from the SIDM component in the galaxy subsequently affects the baryonic disk, since the deepened DM potential increases the gas density and boosts star formation. Thus the long-term co-evolution of the SIDM and baryons leads to a slightly higher stellar mass, a slightly more dense DM halo in the disk plane, and slightly higher star formation rates than in CDM. These differences are evident in our Table \ref{tab:gal_prop} and in Fig.~\ref{fig:dens_dm_rad}, and are the means by which SIDM solves the so-called ``diversity problem.'' 

At the Solar Circle, the distribution and properties of baryonic matter play a role equal to or larger than DM's in creating the observed Galactic acceleration profiles (see Fig.~\ref{fig:m12f_time}). The steeper vertical acceleration gradients in SIDM relative to the same halo in CDM (Fig.~\ref{fig:da_dz}), which are calculated at the Solar Circle, thus reflect the response of both species to the change in the DM model. However, the variation in the acceleration profile among CDM galaxies, which is simply due to their different star formation and assembly histories, spans nearly as large a range. Thus, observations of the density distribution in the MW alone are unlikely to distinguish CDM from SIDM. However, a large sample of measurements of the disk-plane density of galaxies could potentially show a signal, since in SIDM one would expect the mean density in such a sample to be statistically higher than expected from CDM. The trend of density with radius in such a sample may also help distinguish the two theories, since the differences in density between CDM and SIDM become more pronounced at smaller radius (where one expects more frequent self-interactions). At the Solar Circle (8.1 kpc), with a cross-section of 1 cm$^{2}$ g$^{-1}$, we expect about 1 scattering event per Hubble time ($\sim 10$ Gyr) per DM particle in our simulations, with much higher scattering rates in the inner regions of the galaxy \citep{vargya2022shapes}.

Measuring Galactic accelerations may help to illuminate the merger history of the MW, since mergers induce asymmetries in the acceleration profile \citep{chakrabarti2021measurement} (see Fig.~\ref{fig:da_dz},~\ref{fig:m12f_time} in this paper). This highlights the need for flexible disequilibrium models to fully utilize acceleration data \citep{donlon2024galactic}, given that merger-induced variations often exceed the differences induced by changing the DM model. Additionally, we only explored the Galactic acceleration profiles in simulations run with standard Monte Carlo implementation for SIDM \citep{rocha2013cosmological, peter2013cosmological} using the FIRE-2 prescription for baryons \citep{hopkins2018}. In the future, one should explore different simulation suites performed with varied baryon prescriptions \citep[e.g.][]{teyssier2002cosmological, menon2015adaptive, wadsley2017gasoline2, weinberger2020arepo}  and alternative SIDM implementations \citep[e.g.][]{vogelsberger2012subhaloes, fry2015all, meskhidze2022comparing}.

Our main conclusions are summarized below:
\begin{enumerate}
    \item The Solar Neighborhood in the MW is comparatively denser than the median of azimuthally averaged Solar Circle regions in these three FIRE-2 simulations with CDM and SIDM model. (see Sec.~\ref{sec:solar_neigh} and Table~\ref{tab:gal_prop}). Regions with local density similar to the Solar Neighborhood are relatively limited and make up about 6\% of the volume at Solar Circle. The higher density in the MW can be attributed to the relatively thin MW disk even compared to other $\sim 10^{12}$ \Msol{} galaxies \citep{mccluskey2024disc} and/or our placement in a dense region of the Galaxy. It should be noted that we did not explore all of the FIRE-2 MW-mass simulations, so a systematic comparison to better quantify this in relation to the MW remains to be conducted.   

    \item The shape of the potential and density in the Solar Neighborhood are predominantly influenced by the Galactic disk (see Sec.~\ref{sec:oblt}). However, SIDM particles show greater responsiveness to the disk potential leads to higher flattening in the DM shape at Solar Circle. This leads to measurable distinctions in the local dark matter density within the Solar Neighborhood. SIDM halos consistently demonstrate denser and more oblate potentials, resulting in vertical acceleration gradient profiles that are steeper by 10-30\% compared to CDM. (see Sec.~\ref{sec:da_dz_mean} and Fig.~\ref{fig:da_dz}). However, the galaxy-to-galaxy variation in density is broader than the difference between SIDM and CDM at 1 cm${}^2$/g. 

    \item Recent mergers with total mass equal to or greater than the Sagittarius dwarf or the SMC will induce measurable asymmetries across the midplane in the vertical acceleration gradient profiles \citep{Chakrabartietal2019, chakrabarti2020toward}, which can persist over extended periods to the present day (see Sec.~\ref{sec:m12f_acc_time} and Fig.~\ref{fig:m12f_time}). These asymmetries---influenced by factors such as satellite mass and orbit---can be larger than the systematic difference between SIDM and CDM, which poses challenges for probing the nature of DM in a single galaxy. Nonetheless, these asymmetric profiles are a promising probe of the merging history of a galaxy.
\end{enumerate}

\begin{acknowledgments}
AA and RES acknowledge support from the Research Corporation through the Scialog Fellows program on Time Domain Astronomy, from NSF grant AST-2007232, and from NASA grant 19-ATP19-0068. RES is supported in part by a Sloan Fellowship. SC acknowledges support from Research Corporation through the Scialog Fellows program on Time Domain Astronomy, and from NSF AST 2009828.  AW received support from: NSF via CAREER award AST-2045928 and grant AST-2107772; NASA ATP grant 80NSSC20K0513; HST grant GO-16273 from STScI. SL acknowledges support from NSF grant AST-2109234 and HST grant AR-16624 from STScI. LN acknowledges support of NSF through the CAREER award AST-2337864 and grant AST-2307788 as well as the Sloan Foundation.   

This research is part of the Frontera computing project at the Texas Advanced Computing Center (TACC). Frontera is made possible by National Science Foundation award OAC-1818253. Simulations in this project were run using Early Science Allocation 1923870, and analyzed using computing resources supported by the Scientific Computing Core at the Flatiron Institute. This work used additional computational resources of the University of Texas at Austin and TACC, the NASA Advanced Supercomputing (NAS) Division and the NASA Center for Climate Simulation (NCCS), and the Extreme Science and Engineering Discovery Environment (XSEDE), which is supported by National Science Foundation grant number OCI-1053575.
\end{acknowledgments}

\software{IPython \citep{ipython}, Matplotlib \citep{matplotlib}, Numpy \citep{numpy}, Scipy \citep{scipy}, \texttt{halo\_analysis} \citep{2020ascl.soft02014W},
\texttt{gizmo\_analysis} \citep{2020ascl.soft02015W}, CMasher \citep{cmasher}}

\bibliography{main}{}
\bibliographystyle{aasjournal}

\end{document}